\documentclass[12pt]{article}
\usepackage{putex}

\usepackage{cancel}
\usepackage{simplewick}
\usepackage{multirow}
\usepackage{comment}
\usepackage{bbold}
\usepackage{graphicx}
\usepackage{caption}
\usepackage{amsmath}
\usepackage{array}
\usepackage{subcaption}
\usepackage{epstopdf}
\usepackage{enumerate}
\usepackage{cite}
\usepackage{youngtab}
\usepackage{tensor}
\usepackage{slashed}
\usepackage[aligntableaux=center]{ytableau}
\usepackage[utf8]{inputenc}
\usepackage{rotating}
\usepackage{bigfoot}

\usepackage[
colorlinks=true,
linkcolor=black,
urlcolor=blue,
filecolor=black,
citecolor=red,
]{hyperref}

\usepackage[compat=1.1.0]{tikz-feynman}
\usetikzlibrary{decorations.markings}

\numberwithin{equation}{section}

\def \( {\left(}
\def \) {\right)}
\def \< {\left<}
\def \> {\right>}

\newcommand{\be}{\begin{equation}} \newcommand{\ee}{\end{equation}}
\newcommand{\bea}{\begin{eqnarray}}  \newcommand{\eea}{\end{eqnarray}}
\newcommand{\nn}{\nonumber}

\newcommand{\bq}{{\bf q}}

\tikzfeynmanset{
	O/.style = {very thick}
}
\tikzfeynmanset{
	Ot/.style = {scalar}
}
\tikzfeynmanset{
	Th/.style = {double distance=2pt}
}
\tikzfeynmanset{
	dO/.style = {fermion}
}

\begin{document}

	\vspace*{-1.5cm}
	\begin{flushright}    % Publication numbers
		{\small
		}
	\end{flushright}
	
	\vspace{1.8cm}
	\begin{center}        % Main title
		\Huge Convexity of Charged Operators in CFTs with Multiple Abelian Symmetries
	\end{center}
	
	\vspace{0.7cm}
	\begin{center}        % Authors
		{\large  Eran Palti$^1$ and Adar Sharon$^{2}$}
	\end{center}
	
	\vspace{0.15cm}
	\begin{center}        % Institutes
		\emph{$^1$ Department of Physics, Ben-Gurion University of the Negev, Be'er-Sheva 84105, Israel}\\[.3cm]
		\emph{$^2$ Department of Particle Physics and Astrophysics, Weizmann Institute of Science, \\ Rehovot 7610001, Israel}\\[.4cm]
		
		e-mail: \tt palti@bgu.ac.il, adar.sharon@weizmann.ac.il
	\end{center}
	
	\vspace{1.5cm}
	
	%%%%%%%%%%%%%%%%%%%%%%%%%%%%%%%%%%%%%%%%%%%%%%%
	%%%%%%%%%%%%%%%%%%%%%%%%%%%%%%%%%%%%%%%%%%%%%%%
	%%%%%%%%%%%%%%%%%%%%%%%%%%%%%%%%%%%%%%%%%%%%%%%
	%%%%%%%%%%%%%%%%%%%%%%%%%%%%%%%%%%%%%%%%%%%%%%%
	%%%%%%%%%%%%%%%%%%%%%%%%%%%%%%%%%%%%%%%%%%%%%%%
	%%%%%%%%%%%%%%%%%%%%%%%%%%%%%%%%%%%%%%%%%%%%%%%
	%%%%%%%%%%%%%%%%%%%%%%%%%%%%%%%%%%%%%%%%%%%%%%%
	%%%%%%%%%%%%%%%%%%%%%%%%%%%%%%%%%%%%%%%%%%%%%%%
	
	\begin{abstract}
		\noindent Motivated by the Weak Gravity Conjecture in the context of holography in AdS, it has been proposed that operators charged under global symmetries in CFTs, in three dimensions or higher, should satisfy certain convexity properties on their spectrum. A key element of this proposal is the charge at which convexity must appear, which was proposed to never be parametrically large. In this paper, we develop this constraint in the context of multiple Abelian global symmetries. We propose the statement that the convex directions in the multi-dimensional charge space should generate a sub-lattice of the total lattice of charged operators, such that the index of this sub-lattice cannot be made parametrically large. In the special case of two-dimensional CFTs, the index can be made parametrically large, which we prove by an explicit example. However, we also prove that in two dimensions there always exist convex directions generating a sub-lattice with an index bounded by the current levels of the global symmetry. Therefore, in two dimensions, the conjecture should be slightly modified to account for the current levels, and then it can be proven. In more than two dimensions, we show that the index of the sub-lattice generated by marginally convex charge vectors associated to BPS operators only, can be made parametrically large. However, we do not find evidence for parametric delay in convexity once all operators are considered.
	\end{abstract}
	
	\thispagestyle{empty}
	\clearpage
	
	\tableofcontents
	
	\setcounter{page}{1}
	
	%%%%%%%%%%%%%%%%%%%%%%%%%%%%%%%%%%%%%%%%%%%%%%%
	%%%%%%%%%%%%%%%%%%%%%%%%%%%%%%%%%%%%%%%%%%%%%%%
	%%%%%%%%%%%                 %%%%%%%%%%%%%%%%%%%
	%%%%%%%%%%%  DOCUMENT BODY  %%%%%%%%%%%%%%%%%%%
	%%%%%%%%%%%                 %%%%%%%%%%%%%%%%%%%
	%%%%%%%%%%%%%%%%%%%%%%%%%%%%%%%%%%%%%%%%%%%%%%%
	%%%%%%%%%%%%%%%%%%%%%%%%%%%%%%%%%%%%%%%%%%%%%%%
	%%%%%%%%%%%%%%%%%%%%%%%%%%%%%%%%%%%%%%%%%%%%%%%
	
	%\newpage
	
	%%%%%%%%%%%%%%%%%%%%%%%%%%%%%%%%%%%%%%%%%%%%%%%
	\section{Introduction}
	\label{sec:intro}
	%%%%%%%%%%%%%%%%%%%%%%%%%%%%%%%%%%%%%%%%%%%%%%%
	
	Motivated by the Weak Gravity Conjecture \cite{Arkani-Hamed:2006emk} (see \cite{Palti:2019pca,Palti:2020mwc,Harlow:2022gzl} for reviews), or more precisely the closely connected Repulsive Force Conjecture \cite{Palti:2017elp,Heidenreich:2019zkl}, a number of conjectures were made in \cite{Aharony:2021mpc} regarding quantum gravity in anti-de Sitter space and about Conformal Field Theories (CFTs). In particular, it was proposed that in a unitary CFT, in $d \geq 3$ dimensions, which has a $U(1)$ global symmetry, the spectrum of charged operators must satisfy a certain convexity property. The Abelian Convex Charge conjecture is a bound
	\be
	\Delta \left(n_1q_0 + n_2q_0\right) \geq \Delta \left( n_1 q_0\right) + \Delta \left( n_2 q_0\right) \;,
	\label{accc}
	\ee
	where $\Delta\left(q\right)$ denotes the dimension of the lowest dimension operator of charge $q$ under the global $U(1)$ symmetry, $n_1$ and $n_2$ are any positive integers, and $q_0$ is an integer charge. A specific generalization of the bound was also proposed for the non-Abelian global symmetry case. The conjecture was tested in a number of CFTs in \cite{Aharony:2021mpc,Dupuis:2021flq,Antipin:2021rsh,Moser:2021bes}, and so far is satisfied. It was also studied further in \cite{Aalsma:2021qga,Watanabe:2022htq}.
	
	The charge parameter $q_0$ is a central aspect of the conjecture. Indeed, for sufficiently large $q_0$, one expects to enter the large charge regime of CFTs where the dimension of charged operators behaves as $\Delta\left(q\right) \sim q^{\frac{d}{d-1}}$ \cite{Hellerman:2015nra}, and so the spectrum is convex. Therefore, much of the content of the conjecture is held in what $q_0$ is. In \cite{Aharony:2021mpc}, it was proposed that $q_0$ cannot be made parametrically large in any parameter of the CFT. In this paper, we study aspects of this statement. 
	
	In most cases of CFTs we are led to think about multiple $U(1)$ symmetries, and this requires a formulation of what is the condition on convexity that we should demand in such settings. The convexity property (\ref{accc}) is a one-dimensional statement, in the sense that for multiple $U(1)$s we can replace the charge $q_0$ by some vector of charges ${\bf q}$, and then demand convexity for multiples of that vector. More precisely: consider a CFT which has $M$ $U(1)$ global symmetries. The charge of an operator is then given by an $M$-component integer vector $\bq$. A {\it convex vector} $\bq$ then satisfies
	\be
	\Delta\left(\left(n_1+n_2\right)\bq\right) \geq \Delta\left(n_1 \bq \right) + \Delta\left(n_2 \bq \right) \;,
	\label{mu1cc}
	\ee
	for any positive integers $n_1$ and $n_2$.
	In terms of the motivation from holography, as in \cite{Aharony:2021mpc}, this can be motivated by a bulk particle charged under the dual gauge $U(1)$s with charge $\bq$ having a positive self-binding energy.
	
	 But how many such convex vectors should we demand within a higher dimensional charge space? This is not clear. In \cite{Aharony:2021mpc}, only a mild statement was made, that for $M$ $U(1)$s there should be $M$ independent convex directions in charge space. Another question which arises in this context is what replaces the condition that $q_0$ cannot be parametrically large in the case of multi-dimensional charge spaces? To make a sharp condition on convexity to test, for multiple $U(1)$ global symmetries, we propose the following:
	\newline
	
	{\bf Convexity Index Conjecture:} {\it There should exist a set of convex charge vectors ${\bf q}_i$, such that this set defines a basis for a sub-lattice of the full charge lattice, which has a sub-lattice index $I$ that is finite and cannot be made parametrically large}.
	\newline
	
	Note that we are not making a statement about a notion of convexity for the sub-lattice here. We are using the index of the sub-lattice to make a statement only about the convexity of the basis vectors themselves, not the full lattice they generate.
	
	We test the conjecture in various CFTs. In two dimensions, we show through an explicit example that it is violated in that the index $I$ can be made parametrically large. On the other hand, we prove that the index is bounded by the levels of the extended chiral algebras associated to the $U(1)$ symmetries. So that while it can be made parametrically large, this is bounded in a precise and specific way, which is specified as part of the data associated to the currents. So, in two dimensions, the conjecture must be slightly refined such that the index $I$ cannot be made larger than the levels of the $U(1)$ currents, and then it can be proven.
	
	In three dimensions or higher, we do not find any examples where the index $I$ can be shown to be parametrically large. In particular, attempts to implement a mechanism similar to the one which lead to a parametrically large index in two dimensions do not seem to work in three or higher dimensions. We do show that in supersymmetric theories the index of the sub-lattice generated by marginally convex charge vectors associated to BPS operators only can be made parametrically large. Therefore, in such theories, if convexity with an order-one index manifests, it must be through the non-BPS operator spectrum.
	
	%%%%%%%%%%%%%%%%%%%%%%%%%%%%%%%%%%%%%%%%%%%%%%%
	\section{Lattice index bound}
	\label{sec:lattind}
	%%%%%%%%%%%%%%%%%%%%%%%%%%%%%%%%%%%%%%%%%%%%%%%
	
	Having introduced the Convexity Index Conjecture, in this section we develop the motivation for it, and suggest also a stronger version of it.
	
	Given a convex vector (\ref{mu1cc}), we are interested in imposing something similar to what we imposed in the single $U(1)$ case: that $q_0$ is of order one, or more precisely not parametrically large. We can consider asking that $\bq$ has components which are of order one. However, this is a basis-dependent condition. For example, consider the case of two $U(1)$s, which we label $U(1)_1$ and $U(1)_2$. Then we can consider a state with charge under these of $\left(q_1,q_2\right)=\left(1,0\right)$. But now we can perform a unimodular transformation to a different $U(1)$ basis
	\be
	U(1)'_1 = \left(N+1\right)U(1)_1 - U(1)_2 \;,\;\; U(1)'_2 = -N U(1)_1 + U(1)_2 \;,
	\label{unimodtra}
	\ee 
	where $N$ is any integer. Both bases are good bases for the same lattice of charged operators. The charge of the same operator in the new basis $U(1)'_1$ and $U(1)'_2$ is $\left(q'_1,q'_2\right)=\left(N+1,-N\right)$.  This appears to be parametrically large.  We therefore would like to make a statement which is basis-independent. 
	
	Another issue arises from guidance from the Weak Gravity Conjecture in the bulk. In the case of multiple $U(1)$ symmetries, the condition which is typically imposed is that there should exist particles whose convex hull in the $g q / m$ plane includes the unit ball \cite{Cheung:2014vva}, where $g$ is the gauge coupling of the $U(1)$, $q$ the charge under it, and $m$ the mass of the particle. This condition allows for the decay of charged black holes (in pure Einstein-Maxwell theories). If we consider charge vectors in the $q$ plane, they can be very different to vectors in the $gq$ plane. For example, consider a bulk theory with gauge couplings $g_1=1/N,g_2=1$, with $N$ some large integer. In the lattice of charges, the vector $\left(q_1,q_2\right)=\left(N,1\right)$ seems to be aligned almost completely with the $q_1$ axis. But in the plane $\left(g_1 q_1, g_2 q_2 \right)$ it is in the direction $\left(1,1\right)$, which is not aligned with the $g_1 q_1$ direction. In terms of the convex hull, particles with charges $\left(q_1,q_2\right)=\left(1,0\right)$ and $\left(q_1,q_2\right)=\left(N,1\right)$ give vectors whose directions are not aligned in the $g q$ plane, and so their convex hull can include the unit ball. While purely from the charge lattice perspective, their directions would be extremely aligned.  
	
	We therefore seek a measure for convexity which is independent of the basis used to describe the charge lattice of operators, and which also would give the same results in the $q$ and $gq$ planes, thus avoiding the ambiguity of which plane to use. There is a very natural such measure that can be constructed as follows. Take some subset of the convex directions in the lattice of charged operators. Let us denote them as $\bq_i$. Then these vectors can act as a basis for a sub-lattice of charged operators within the full lattice. This sub-lattice has an index associated to it, which corresponds to the volume of a sub-lattice cell in units of the volume of the full lattice cell. The conjecture is then that there exists some choice of convex directions which gives such an index that is non-vanishing and not parametrically large.
		
	Calculating the index is not always simple. Let us assume that the charge lattice of operators spans the usual full integer lattice, so the set of vectors with integer entries. Then there is a simple way to calculate the index: form a square matrix from $M$ of the $\bq_i$, and then the index is the determinant of this matrix. It is more convenient to work with a normalized index $I$, so we consider the $M^{\mathrm{th}}$ root of the (absolute value of the) determinant of this matrix
	\be
	I = \big| \mathrm{det\;} \left(\bq_1\;\bq_2 \;...\;\bq_M\right) \big|^{\frac{1}{M}}\;.
	\label{index}
	\ee
	
In the case of a sub-lattice basis which is orthogonal, the index matches the simplest natural extension of $q_0$ to two lattice directions, giving the geometric average of the minimal charge along each direction. So say we have convex vectors $\left(q_0^1,0\right)$ and $\left(0,q_0^2\right)$, then the index is $I=\sqrt{q_0^1 q_0^2}$. Of course, the sub-lattice index is independent of the basis used to describe the sub-lattice. So, for example, we can consider a unimodular transformation of bases as in (\ref{unimodtra}). This would mean that the convex vectors above would now read $\left(\left(N+1\right)q_0^1,-Nq_0^1\right)$ and $\left(-q_0^2,q_0^2 \right)$, whose determinant still gives the same index.
	
	When the convex vectors are not orthogonal, the index can give perhaps unintuitive results. For example, consider two convex vectors $\left(1,0\right)$ and $\left(N,1\right)$, for large $N$. This might seem like a highly non-convex situation, since one of the vectors has a large component, but the index is $I=1$. The corresponding sub-lattice is very elongated, with a cell that is much longer than it is tall, and with a volume that is not large. 
	
	This volume picture also addresses the issue of the normalization of the currents, or the gauge couplings in the bulk. The lattice of $gq$ is related to the lattice in $q$ through a linear transformation. Linear transformations preserve the sub-lattice index, and so give the same answer if one considers $gq$ or $q$. We do not have the issue of vectors seeming aligned in the $q$ plane, but not being aligned in the $gq$ plane. 
	
	So far we assumed that the lattice of charged operators spans the integer lattice in $\mathbb{R}^M$, so the set of vectors with integer entries. Typically, the operator charges are not distributed so simply. However, given a basis of charges for operators, we can always map it to the unit integer basis by some linear transformation. Since the index is invariant under linear transformations acting on the lattice, the calculation of the index can then be done using the determinant formula (\ref{index}) for the transformed sub-lattice vectors. 
	
	Let us note that there is no sense in which the sub-lattice generated by the convex vectors is a convex lattice. Convexity is a property along vectors, not of a lattice. We are only using the index of a sub-lattice to quantify naturally the magnitude of the convex vectors. 
	
	However, in flat space, there is a version of the Weak Gravity Conjecture termed the sub-lattice WGC \cite{Heidenreich:2016aqi}, which motivates a strong version of the Convexity Index Conjecture. The Sub-lattice WGC is a statement about the spectrum of single-particle states (or some long-lived bound states) in the theory, and really asks that a lattice is populated. It is therefore very different to the sub-lattice used in the initial Convexity Index Conjecture, where there are no statements made about the lattice but only about the existence of appropriate basis vectors. Further, the lattice of operators in the CFT can be generated by a single bulk particle, so there is no requirement for multiple particles. The two ideas are therefore distinct. A statement in the CFT which is more closely related to the sub-Lattice WGC, or more precisely some combination of its flat space repulsive force version \cite{Heidenreich:2019zkl} and the Positive Binding Conjecture in AdS of \cite{Aharony:2021mpc}, is a stronger version on the Convexity Index Conjecture:
	\newline
	
	{\bf Strong Convexity Index Conjecture:} {\it The sub-lattice defined by the basis of convex charge vectors ${\bf q}_i$ is further constrained to be such that each point in it is itself a convex vector (relative to the origin). The index of this sub-lattice is also finite and cannot be made parametrically large.}
	\newline
	
	This is clearly a much stronger statement than the weaker version of the conjecture, turning it into a statement about the existence of an infinite number of convex vectors (one for each direction in the lattice). It therefore should be considered with more caution. Still, we do not find any counter-example to this strong version in the theories studied in this paper. Further, the proof of the weaker version in two-dimensional CFTs holds equally for the strong version.
		
	In \cite{Aharony:2021mpc}, a general convexity conjecture was proposed which included non-Abelian global symmetries. In the case of non-Abelian symmetries, the basis is not an issue since they cannot mix with the Abelian symmetries or with each other. In this sense they are simpler to consider. The most natural way to include non-Abelian symmetries in the index constraint is to take their Cartan sub-algebra as forming part of the $U(1)$ lattice. The operators transforming in representations of the non-Abelian group then lead to charges given by their weights. This matches the non-Abelian conjecture in \cite{Aharony:2021mpc} which proposed the existence of representations with order one weights. We propose that one only considers a single $U(1)$ representative from the Cartan sub-algebra of each non-Abelian factor. 
	
	%%%%%%%%%%%%%%%%%%%%%%%%%%%%%%%%%%%%%%%%%%%%%%%
	\section{Two-dimensional CFTs}
	\label{sec:2dcfts}
	%%%%%%%%%%%%%%%%%%%%%%%%%%%%%%%%%%%%%%%%%%%%%%%
	
	In \cite{Aharony:2021mpc}, the convexity conjecture was proposed for CFTs in three or more dimensions. This is natural in the sense that holography with a three-dimensional bulk, and two-dimensional CFT, is rather special. There are no propagating gravitational degrees of freedom in the bulk, and the $U(1)$ global currents in the CFT are dual to bulk Chern-Simons terms (of gauge fields with boundary-localised degrees of freedom). Indeed, there is no natural reason to expect something like the Weak Gravity Conjecture to hold in three dimensions. For a nice discussion of these points we refer to \cite{Montero:2016tif}. Further, a possible counter example, in which one could consider parametrically delaying convexity was raised. In this section we will study this counter example in detail, applying the newly introduced measure for convexity. 
	
	In fact, two-dimensional CFTs are an excellent setting to study convexity properties because, as we show below, we can find an upper bound on the index, so the sense in which one can parametrically delay convexity can be made very precise. Therefore, rather than dismissing two dimensional CFTs, we will show that they require a precise modification of the conjecture, and after this modification, they can be rigorously proven. 
	
	%%%%%%%%%%%%%%%%%%%%%%%%%%%%%%%%%%%%%%%%%%%%%%%
	\subsection{Bound on the convexity index}
	\label{sec:ibou}
	%%%%%%%%%%%%%%%%%%%%%%%%%%%%%%%%%%%%%%%%%%%%%%%
	
	Consider a two-dimensional CFT with a global $U(1)$ symmetry and associated conserved current $j_\mu$. Let us split the current into holomorphic $J$ and anti-holomorphic $\bar{J}$ pieces
	\be
	j_{0}\left(z,\bar{z}\right) = J\left(z\right) + \bar{J}\left(\bar{z}\right)\;, \; j_{1}\left(z,\bar{z}\right) = J\left(z\right) - \bar{J}\left(\bar{z}\right)  \;.
	\ee
%	
%	
%	A global $U(1)$ symmetry is associated to a holomorphic conserved current $J\left(z\right)$. We will assume also an anti-holomorphic current $\bar{J}\left(\bar{z}\right)$, and so a single $U(1)$ also implies a second $U(1)$.\footnote{It is possible to have only a holomorphic current, and $\bar{J}=0$. In such a case, the discussion in this section follows unchanged, setting $\bar{k}=0$, since it focuses only on the holomorphic part.} 
	The holomorphic current combines with the Virasoro generators $L_m$ into an extended chiral algebra. Specifically, expanding 
	\be
	J\left(z\right) = \sum_n \tilde{j}_n z^{-\left(n+1\right)} \;, 
	\ee
	we have the commutation relations
	\be
	\left[\tilde{j}_m,\tilde{j}_n\right] = k \delta_{m+n,0} \;,\;\; \left[L_m,\tilde{j}_n\right]=-n \tilde{j}_{n+m} \;.
	\ee
	$k$ is known as the level of the extended chiral algebra. It also appears in the OPE for the currents
	\be
	J\left(z\right) J\left(0\right) \sim \frac{k}{z^2} \;.
	\label{hollevel}
	\ee
	If the extended chiral algebra is a $U(1)$ algebra, then $k$ must be an integer and we denote the algebra as $U(1)_k$.\footnote{Note that we have assumed that the currents $J$ and $\bar{J}$ are correctly normalized to give integer charges, which fixes the values of $k$ and $\bar{k}$. See, for example, \cite{Lin:2019kpn,Benjamin:2020swg} for a discussion on this normalization.} If the algebra is instead $\mathbb{R}$, $k$ is not necessarily an integer, and in particular it can be irrational. 
	%For simplicity, let us initially consider the case of a $U(1)_k$ algebra where $k$ is an integer, and return to the irrational case in section \ref{sec:knonint}.  
	
	There is similarly an anti-holomorphic level $\bar{k}$ associated to the anti-holomorphic current
	\be
	\bar{J}\left(\bar{z}\right) \bar{J}\left(0\right) \sim \frac{\bar{k}}{\bar{z}^2} \;.
	\label{hollevel}
	\ee
	In complete generality, it is possible to have $\bar{J} = \bar{k} = 0$. In such cases, the analysis follows similarly, but with only the holomorphic part. We will assume though that $k$ and $\bar{k}$ are non-vanishing. This means that we actually have two $U(1)$ symmetries, with extended chiral and anti-chiral algebras $U(1)_k \times U(1)_{\bar{k}}$. 
	
	We would like to consider charged operators. We denote their charges under $J$ and $\bar{J}$ as $q$ and $\bar{q}$ respectively. 	Let us also denote the dimension of a charged operator as $\Delta$, and split this into holomorphic $h$ and anti-holomorphic $\bar{h}$ pieces\footnote{More precisely, our dimension is the difference between the dimension of the operator and the dimension of the vacuum.}
	\be
	\Delta\left(q,\bar{q}\right) = h\left(q\right) + \bar{h}\left(\bar{q}\right) \;.
	\ee
	A useful fact for us is that charged operators in two-dimensional CFTs satisfy a unitarity bound on their holomorphic and anti-holomorphic dimensions by their charge 
	\be
	h\left(q\right) \geq \frac{q^2}{2k}\;, \;\; \bar{h}\left(\bar{q}\right) \geq \frac{\bar{q}^2}{2\bar{k}}\;.
	\label{unb}
	\ee
	This follows from the Sugawara construction of a component of the energy-momentum tensor as a product of currents. See, for example, \cite{Kraus:2006wn,Montero:2016tif}. 
	
	CFTs also have a property called Spectral Flow, which relates operators of different dimensions and charges under $U(1)$ symmetries. See \cite{Lin:2019kpn,Benjamin:2020swg} for a good account. In the context of holography, this first featured in \cite{Maldacena:2000hw}, and more recently was used also in the context of the Weak Gravity Conjecture in \cite{Montero:2016tif,Lin:2019kpn}.\footnote{It was also used in the context of the WGC, but not related to holography, in \cite{Heidenreich:2016aqi}.} Spectral flow is performed along a $U(1)$ direction. Let us label the holomorphic and anti-holomorphic charges under the $U(1)$ as $q$ and $\bar{q}$ (which need not be integer). Spectral flow implies that if the theory has an operator with dimensions $h$ and $\bar{h}$, then there must also exist an infinite set of operators with dimensions $h_m$, $\bar{h}_{m}$ and charges $q_m$, $\bar{q}_{m}$ given by
	\bea
	q_m &=& q - m k \;,\;\; h_m = h - m q + \frac{k m^2}{2} = \frac{q_m^2}{2k} + \left(h - \frac{q^2}{2k} \right) \;, \nn \\
	\bar{q}_{m} &=& \bar{q} + m \bar{k} \;,\;\; \bar{h}_{m} = \bar{h} + m \bar{q} + \frac{\bar{k} m^2}{2} = \frac{\bar{q}_{m}^2}{2\bar{k}} + \left(\bar{h} - \frac{\bar{q}^2}{2\bar{k}} \right)  \;,
	\label{specflow}
	\eea
	where $m$ is an arbitrary integer. An important property of spectral flow is that it maintains saturation of the unitarity bound (\ref{unb}). So if the initial state saturates the bound, so do the the states that it flows to.

Since in our setting we have two $U(1)$ symmetries, we have two spectral flows.  For clarity, let us denote the two $U(1)$ symmetries as $U(1)_n \times U(1)_w$. Their associated (non-holomorphic) currents are $j_n(z,\bar z)$ and $j_w(z,\bar z)$, and these yield integer charges in the OPE with charged operators. These are given in terms of combinations of holomorphic and anti-holomorphic pieces $J(z),\bar J(\bar z)$; in components we can write
	\bea
	j_{n,0}\left(z,\bar{z}\right) = J_n\left(z\right) + \bar{J}_n\left(\bar{z}\right)\;, &\;& j_{n,1}\left(z,\bar{z}\right) = J_n\left(z\right) - \bar{J}_n\left(\bar{z}\right)  \;,\nn \\
	j_{w,0}\left(z,\bar{z}\right) = J_w\left(z\right) + \bar{J}_w\left(\bar{z}\right)\;, &\;& j_{w,1}\left(z,\bar{z}\right) = J_w\left(z\right) - \bar{J}_w\left(\bar{z}\right) \label{U1tohol} \;.
	\eea
	There is only one holomorphic conserved current, and one anti-holomorphic one, and therefore there must be a linear relation between the respective pieces
	\be
	J_w\left(z\right) = \alpha J_n\left(z\right) \;,\;\; \bar{J}_n\left(\bar{z}\right) = \bar{\alpha} \bar{J}_w\left(\bar{z}\right) \;,
	\ee
	where $\alpha$ and $\bar{\alpha}$ are some constants.
	We can then relate the currents to each other
	\be
	j_{n,0} = \frac{J_w}{\alpha} + \bar{\alpha} \bar{J}_w \;,\;\; j_{w,0} = \alpha J_n + \frac{\bar{J}_n}{\bar{\alpha}} \;.
	\ee
	Spectral flow along the $U(1)$s acts simultaneously on the holomorphic and anti-holomorphic parts. So applying (\ref{specflow}) for a flow along $U(1)_w$ by an integer parameter $m_{(w)}$ changes the charge under $U(1)_n$, denoted by $q^{(n)}$, as
	\be
	q^{(n)}_{m_{(w)}} = q^{(n)} + m_{(w)}\left(-\frac{k_w}{\alpha} + \bar{\alpha} \bar{k}_w \right) \;,
	\ee
	and the charge under $U(1)_w$ as
		\be
	q^{(w)}_{m_{(w)}} = q^{(w)} + m_{(w)}\left(-k_w + \bar{k}_w \right) \;.
	\ee
	Similarly, performing flow along $U(1)_n$ by $m_{(n)}$ changes the charges according to
	\be \label{eq:spectral_flow_n}
	q^{(n)}_{m_{(n)}} = q^{(n)} + m_{(n)}\left(-k_n+ \bar{k}_n \right)\;,\qquad 
	q^{(w)}_{m_{(n)}} = q^{(w)} + m_{(n)}\left(- \alpha k_n + \frac{\bar{k}_n}{\bar{\alpha}} \right) \;.
	\ee
	It is worth noting some things about these expression. First, charge quantization demands
	\be
	k_n - \bar{k}_n \in \mathbb{Z} \;,\;\; - \alpha k_n + \frac{\bar{k}_n}{\bar{\alpha}} \in \mathbb{Z} \;,\;\; k_w - \bar{k}_w \in \mathbb{Z}  \;,
	\ee
And similar conditions using the relations $k_w=\alpha^2 k_n,\;\bar{k}_w=\frac{1}{\bar{\alpha}^2}\bar{k}_n$. Second, a $U(1)$ spectral flow only modifies its own charges if the $U(1)$ is anomalous, that is $k \neq \bar{k}$. A special case of this is $\bar{k}=0$, in which case $k$ must be integer and we can then have only a single $U(1)$.
	
	By acting on the vacuum state with $q^{(n)}=q^{(w)}=0$ and $\Delta=0$, we can find a two-parameter lattice of operators which have charges 	
	\be
	\bq_{m_{(w)},m_{(n)}} \equiv \left( \begin{array}{c} q^{(n)}_{m_{(w)},m_{(n)}} \\ q^{(w)}_{m_{(w)},m_{(n)}} \end{array} \right) = m_{(n)}\left( \begin{array}{c}-k_n+ \bar{k}_n  \\
	- \alpha k_n + \frac{\bar{k}_n}{\bar{\alpha}}
	 \end{array} \right) 
	 +m_{(w)} \left( \begin{array}{c}  -\frac{k_w}{\alpha} + \bar{\alpha} \bar{k}_w \\
	 -k_w + \bar{k}_w 
	 \end{array} \right)
	 \ee
	Since the vacuum state saturates the unitarity bound (\ref{unb}), so do all the operators reached from it by spectral flow. Therefore, this lattice of charged operators have minimal dimension for their charge, and therefore are the appropriate operators for the convexity conjecture. 
	
	We can then read off the corresponding spectral flow index
	\be
	I_{SF}=\left|k_{w}\bar{k}_{n}\left(1-\frac{1}{\bar{\alpha}\alpha}\right)+\bar{k}_{w}k_{n}\left(1-\bar{\alpha}\alpha\right)\right|^{1/2}\;.
	\label{sfb2u1_2}
	\ee 
	%\be
	%I_{\mathrm{SF}} = \left|k_n k_w - \alpha \bar{\alpha} k_n \bar{k}_w - \frac{\bar{k}_n k_w}{\alpha\bar{\alpha}} + \bar{k}_n \bar{k}_w\right|^{\frac12}\;,
	%\label{sfb2u1_2}
	%\ee 
%	Alternatively, using $k_{w}=\alpha^{2}k_{n}$ and $\bar{k}_{w}=\frac{\bar{k}_{n}}{\bar{\alpha}^{2}}$ we can write this as
%	\be
%	I_{SF}=\left|k_n\bar{k}_n\right|^{1/2}\left|\frac{1}{\bar{\alpha}}-\alpha\right|
%	=\left|k_w\bar{k}_w\right|^{1/2}\left|\frac{1}{\alpha}-\bar{\alpha}\right|\;.
%	\label{sfb2u1_2}
%	\ee
	We therefore see that the spectral flow index is determined by the levels of the extended chiral algebras.\footnote{Note that there is a special case of having a single holomorphic $U(1)$ with $\bar{k}=0$, which leads to the index $I_{SF}=k$.}
	The index value (\ref{sfb2u1_2}) places an upper bound on the convexity index for the CFT 
	\be
	I \leq I_{\mathrm{SF}} \;.
	\label{indb2u1_2}
	\ee 
	It is an upper bound because there may be convex directions in charge space, not those generated by spectral flow, which lead to a lower index. 
	
	Note that this implies that both the weak and strong versions of the Convexity Index Conjecture are satisfied, up to the refinement that the index can be parametrically large, but bounded by the spectral flow index.
	
	%%%%%%%%%%%%%%%%%%%%%%%%%%%%%%%%%%%%%%%%%%%%%%%
	\subsubsection{Multiple $U(1)$s}
	\label{sec:knonint}
	%%%%%%%%%%%%%%%%%%%%%%%%%%%%%%%%%%%%%%%%%%%%%%%
	
	We have discussed the case of two $U(1)$ symmetries. The generalisation to an arbitrary number of $U(1)$ symmetries follows along the same lines.
	We consider $2M$ $U(1)$s, which we label as $U(1)^a_n$ and $U(1)^b_w$, with $a=1,...,M$. Their associated holomorphic and anti-holomorphic currents have OPEs
	\bea
	J_n^a\left(z\right) J_n^b\left(0\right) &\sim & \frac{k_n^{ab}}{z^2} \;, \;\; \bar{J}_n^a\left(\bar{z}\right) \bar{J}_n^b\left(0\right) \sim \frac{\bar{k}_n^{ab}}{z^2} \;, \nn \\
	J_w^a\left(z\right) J_w^b\left(0\right) &\sim & \frac{k_w^{ab}}{z^2} \;, \;\; \bar{J}_w^a\left(\bar{z}\right) \bar{J}_w^b\left(0\right) \sim \frac{\bar{k}_w^{ab}}{z^2} \;.
	\eea
	The matrices $k^{ab}$ and $\bar{k}^{ab}$ must be positive definite. The partner currents are related as
	\be
	J^a_w\left(z\right) = \alpha^a_{\;\;b} J^b_n\left(z\right) \;,\;\; \bar{J}^a_n\left(\bar{z}\right) = \bar{\alpha}^a_{\;\;b} \bar{J}^b_w\left(\bar{z}\right) \;,
	\ee
	where $\alpha^a_{\;\;b}$ and $\bar{\alpha}^a_{\;\;b}$ are invertible matrices.
	
	Spectral flow is associated to an arbitrary integer vector $\rho_a$, which act on the holomorphic and anti-holomorphic charges as (see, for example \cite{Heidenreich:2016aqi})\footnote{Note that in \cite{Heidenreich:2016aqi} flow was taken with two independent vectors $\rho_a$ and $\bar{\rho}_a$ which had to satisfy a quantization constraint $q^a \rho_a - \bar{q}^a \rho_a \in \mathbb{Z}$. But with the normalization of the $U(1)$s that we take, so that $q^a + \bar{q}^a \in \mathbb{Z}$, and completeness of the spectrum, one should restrict to $\rho_a = - \bar{\rho}_a$, as in \cite{Benjamin:2020swg}.}
	\be
	q^{a}_{m}  = q^{a} - m k^{ab} \rho_b \;,\;\; \bar{q}^{a}_{m}  = \bar{q}^{a} + m \bar{k}^{ab} \rho_b \;.
	\ee
	As in the case with two $U(1)$s, spectral flow preserves the unitarity bound
	\be
	h\left(q\right) \geq \frac{q^a \left(k^{-1}\right)_{ab} q^b}{2} \;, \;\; \bar{h}\left(\bar{q}\right) \geq \frac{\bar{q}^a \left(\bar{k}^{-1}\right)_{ab} \bar{q}^b}{2} \;,
	\ee
	so operators generated by spectral flow from the vacuum will be the lowest dimension for their charge. 
	
	We can then act on the $U(1)$ charges by spectral flow, starting from the vacuum, to find charged operators with charges
	\bea
	q^{(n)}_{m_{(w)},m_{(n)}} &=&  \left[ m_{(w)} \left( - \alpha^{-1} k_w + \bar{\alpha} \bar{k}_w \right) + m_{(n)} \left(-k_n + \bar{k}_n \right) \right] \rho \;,\nn \\
	q^{(w)}_{m_{(w)},m_{(n)}} &=&  \left[m_{(w)} \left( - k_w + \bar{k}_w \right) + m_{(n)} \left(- \alpha k_n + \bar{\alpha}^{-1} \bar{k}_n \right) \right] \rho \;,
	\eea
	where we have suppressed explicit indices. 
	
	There are $M$ spectral flow vectors $\rho_a$, so we can label the set of such vectors as $\left(\rho_a\right)_b$ and form a matrix from them $P_{ab}$. It is then convenient to choose this to be the identity matrix
	\be
	P_{ab} \equiv \left(\rho_a\right)_b = \delta_{ab}\;.
	\ee
	This means that the determinant of the matrix of the sub-lattice vectors associated to the spectral flow charges can be calculated as
	\be
	I_{SF} = \left|\mathrm{det}\; \left(\begin{array}{cc} 
	- \alpha^{-1} k_w + \bar{\alpha} \bar{k}_w & -k_n + \bar{k}_n \\
	- k_w + \bar{k}_w  & - \alpha k_n + \bar{\alpha}^{-1} \bar{k}_n
	\end{array} \right) \right|^{\frac{1}{2M}}\;,
	\label{spigeb}
	\ee
	which is a neat upper bound on the index of any two-dimensional CFT. 
	
Note that we have assumed an equal number of holomorphic and anti-holomorphic currents. Cases with different number of holomorphic and anti-holomorphic currents are simple, since for those additional $U(1)$s the $k$'s must be quantized integers and the $\bar{k}$'s vanish. Each such $U(1)$ then contributes to the index a factor of $k$. If there are multiple such $U(1)$s which mix with a level matrix ${\bf k}$, they will contribute a factor of det ${\bf k}$.	

	%%%%%%%%%%%%%%%%%%%%%%%%%%%%%%%%%%%%%%%%%%%%%%%
	\subsection{Example of a parametrically large index}
	\label{sec:expi}
	%%%%%%%%%%%%%%%%%%%%%%%%%%%%%%%%%%%%%%%%%%%%%%%
	
	We have seen that in two-dimensional CFTs there is a bound on the convexity index by the current levels. The current levels can be made parametrically large. However, this does not imply that the index can be parametrically large, since it is only bounded from above by the levels. In this section we present an explicit example of a CFT which indeed has a parametrically large convexity index.\footnote{This example was first suggested to us by C. Vafa. We also thank O. Aharony for initial joint investigations of it.} We will see that it does not saturate the bound set by the levels, indeed, we did not find any examples which saturate the bound. 
	
	The theory we consider is constructed initially from a product of two theories. The first is the simple theory of a free scalar boson with the target space of a circle. We denote this theory by CFT$_{S^1}$, and we use the conventions in \cite{Benjamin:2020swg}. The theory has a single field $X$, which is periodic $X \sim X + 2 \pi R$, with $R$ the target space circle radius. We can split it into left-moving and right-moving parts $X\left(z,\bar{z}\right) = X_L\left(z\right) + X_R\left(\bar{z}\right)$. The normalisation of the OPEs is
	\be
	\partial X \left(z\right) \partial X\left( 0\right) \sim -\frac{1}{2z^2} \;,\;\;\bar{\partial} X \left(\bar{z}\right) \bar{\partial} X\left( 0\right) \sim -\frac{1}{2\bar{z}^2}\;.
	\ee
	The Virasoro primary operators are
	\be
	{\cal O}_{n,w}\left(z,\bar{z}\right) = \mathrm{exp} \left[i\left(\frac{n}{R} + w R \right) X_L +  i\left(\frac{n}{R} - w R \right) X_R \right] \;, 
	\label{VP}
	\ee 
	and they have the OPEs
	\be
	i \partial X\left(z\right) {\cal O}_{n,w} \left(0\right) \sim \frac{q_L}{z}\; {\cal O}_{n,w}\left(0\right) \;,\;\; i \bar{\partial} X\left(z\right) {\cal O}_{n,w} \left(0\right) \sim \frac{q_R}{\bar{z}}\; {\cal O}_{n,w}\left(0\right) \;,
	\ee
	where
	\be
	q_L = \frac12 \left(\frac{n}{R} + w R \right) \;,\;\; q_R = \frac12 \left(\frac{n}{R} - w R \right) \;.
	\ee
	
	The theory has a global symmetry group $U(1)_n \times U(1)_w$, where $U(1)_n$ is associated to Kaluza-Klein (KK) modes and $U(1)_w$ to winding modes. The conserved currents $j_n,j_w$ are given as in \eqref{U1tohol}, where
	\bea
	J_n=iR\partial X\;, &\;& \bar{J}_n=iR\bar{\partial} X\;, \nn \\
	J_w=\frac{i}{R}\partial X\;, &\;&  \bar{J}_w = -\frac{i}{R}\bar{\partial} X\;.
	\eea
%	\be
%	j_n\left(z,\bar{z}\right) = i R \left( \partial X + \bar{\partial} X \right) \;, \;\; j_w\left(z,\bar{z}\right) = \frac{i}{R} \left( \partial X - \bar{\partial} X \right)  \;,
%	\ee
	The charges under $j_n,j_w$ are associated with $\left(n,w\right)$ denoting the KK and winding numbers, $n,w \in \mathbb{Z}$. The dimensions of the operators (\ref{VP}) are
	\be
	\Delta_{n,w}^{S^1} = q_L^2 + q_R^2 = \frac12 \left( \frac{n^2}{R^2} +  w^2 R^2 \right) \;.
	\ee
	
	The second CFT, which we denote by CFT$_{Y^N}$, is just $N$ copies of any CFT. Let us denote each copy as CFT$_Y$, and denote its central charge as $c_Y$. We now consider the two CFTs together CFT$_S^1 \;\otimes\;$CFT$_{Y^N}$ and in this total theory we gauge a discrete $\mathbb{Z}_N$ symmetry which acts as a cyclic permutation on the $N$ copies of CFT$_Y$ while at the same time acting as a rotation by $\frac{2\pi}{N}$ on the circle in CFT$_{S^1}$. The resulting theory, denoted CFT$_{\mathbb{Z}_N}$, is the theory we will study:
	\be
	\mathrm{CFT}_{\mathbb{Z}_N} = \frac{\mathrm{CFT}_{Y^N} \otimes \mathrm{CFT}_{S^1}}{\mathbb{Z}_N} \;.
	\ee
	
	The theory has the following operators charged under the $U(1)_n \times U(1)_w$ global symmetries. In the winding sector, we are now allowed fractional winding modes which wind only $\frac{2\pi}{N}$ around the circle. We will then change notation such that now $w$ denotes a fractional winding around the circle. However, because the $\mathbb{Z}_N$ action acts also on the $\mathrm{CFT}_{Y^N}$, such fractional winding operators need to be joined to a permutation operator in that sector, which we denote by $\sigma$. The dimension of $\sigma$ is given by (see, for example \cite{Lunin:2000yv})
	\be
	\Delta_{\sigma} = \frac{c_Y}{24}\left(N - \frac{1}{N}\right)\;.
	\ee
	In the KK sector, we still have the KK modes which have momentum that are multiples of $N$ that are left unchanged. If the KK momentum is not a multiple of $N$, then the operator needs to be joined to an operator of the $\mathrm{CFT}_{Y^N}$ which can cancel its $\mathbb{Z}_N$ charge. The appropriate operator, for a KK charge $n$, is 
	\be
	x^n = \sum_j^N \chi_j e^{-\frac{2\pi i j n}{N}} \;,
	\ee 
	where $\chi_j$ is the operator with the lowest dimension inside the $j^{\mathrm{th}}$ copy of the CFT$_Y$ inside $\mathrm{CFT}_{Y^N}$. To see this, note that under a permutation $j\rightarrow j+1$, the operator $x^n$ transforms with a phase $e^{-\frac{2\pi i n}{N}}$, which exactly cancels the phase of the KK mode under the $\mathbb{Z}_N$ rotation. We denote the dimension of $\chi$ as $\Delta_{\chi}$ and take it to be of order one
	\be
	\Delta_{\chi} \sim {\cal O}\left(1\right) \;.
	\ee 
	The full spectrum of operators charged under the two $U(1)$ symmetries is then labelled by KK momentum $n$ and (fractional) winding $w$, and has dimensions
	
	\be
	\Delta_{n,w} =  \frac12\left(\frac{n}{R}\right)^2 +  \frac12\left(\frac{w R}{N}\right)^2  + \Delta_{x\left(n\right)} + \Delta_{\sigma\left(w\right)} \;,
	\ee
	where
	\begin{equation}\label{eq:dimensions}
	\begin{split}
	\Delta_{\sigma\left(w\right)} & = \begin{cases}
	0\;, & \mathrm{if\;} w = 0 \mod N\\
	\Delta_{\sigma}\;, & \text{otherwise}
	\end{cases} \;,\\
	\Delta_{x(n)} & = 
	\begin{cases}
	0\;, & \mathrm{if\;} n = 0 \mod N\\
	\Delta_\chi\;, & \text{otherwise}
	\end{cases}\;.
	\end{split}
	\end{equation}
	
	We can now see why this theory can parametrically delay convexity in the spectrum. Say we consider an operator with non-zero winding number which is not a multiple of $N$. Such an operator can receive an arbitrarily large dimension contribution from the $\Delta_{\sigma\left(w\right)}$ factor by taking an $N$ arbitrarily large. However, its charge remains the same, meaning that its dimension is essentially decoupled from its charge. Such a situation clearly violates convexity (\ref{accc}), since the right hand side will be roughly twice the left hand side. So we only have convexity for winding modes with charge that is a multiple of $N$, thereby parametrically delaying convexity in charge. This is a first impression of why this CFT is an interesting challenge to convexity; we now proceed to perform a more detailed analysis. 
	
	%%%%%%%%%%%%%%%%%%%%%%%%%%%%%%%%%%%%%%%%%%%%%%%
	\subsubsection{Compatibility with the index bound}
	\label{sec:compind}
	%%%%%%%%%%%%%%%%%%%%%%%%%%%%%%%%%%%%%%%%%%%%%%%
	
	Let us check that this CFT is compatible with the general index bound (\ref{indb2u1_2}). To do this we need to work with holomorphic currents rather than those associated to winding and KK modes. Since the OPE of the holomorphic currents is not integer for general R, we should follow the procedure in section \ref{sec:knonint}. First we note that because we now allow fractional winding modes, the appropriately normalized currents are
	\bea
	J_n=iR\partial X\;, &\;& \bar{J}_n=iR\bar{\partial} X\;, \nn \\
	J_w=\frac{iN}{R}\partial X\;, &\;&  \bar{J}_w = -\frac{iN}{R}\bar{\partial} X\;.
	\eea
	The holomorphic parts of the KK and winding currents are related as
	\be
	J_w\left(z\right) = \frac{iN}{R} \partial X = \frac{N}{R^2} \left( iR \partial X \right) = \frac{N}{R^2} J_n\left(z\right) \;.
	\ee
	From this we read that $\alpha = \frac{N}{R^2}$. Similarly, we find $\bar{\alpha}=-\frac{R^2}{N}$.	
	The current levels are extracted from 
	\be
	J_n\left(z\right) J_n\left(0\right) \sim \frac{R^2}{2 z^2} \;,\;\; \bar{J}_w\left(\bar{z}\right) \bar{J}_w\left(0\right) \sim \frac{N^2}{2R^2 \bar{z}^2}  \;,
	\ee
	which gives
	\be
	k_n = \bar{k}_n = \frac{R^2}{2} \;,\;\; k_w = \bar{k}_w = \frac{N^2}{2 R^2}\;.
	\ee
	From (\ref{sfb2u1_2}) we then extract the index bound
	\be
	I \leq I_{\mathrm{SF}} = N \;.
	\ee
	
	To have a convex spectrum we can choose both the KK and winding modes to be multiples of $N$, this essentially negates the $\mathbb{Z}_N$ orbifolding leading to a spectrum of extremal states. This indeed precisely saturates the index bound $I = N$.

	\subsubsection{The minimal index}
	\label{sec:maxind}
	%%%%%%%%%%%%%%%%%%%%%%%%%%%%%%%%%%%%%%%%%%%%%%%
	
	We have seen that choosing the KK and winding modes to be multiples of $N$ gives convex directions in charge space which have an associated convexity index that saturates the bound from spectral flow. However, this index may not be the minimal one, and indeed we show that it is not. It is worth considering two cases separately.
	
	\subsubsection*{$\mathrm{Case\;1}\;:\; R^2 \geq N$}
	
	This is the situation with $R$ large, so if we consider the dimension of operators with KK charges, it is dominated by the contribution from the $\mathrm{CFT}_{Y^N}$ sector. By making $R$ large enough we can always make this contribution arbitrarily larger, and therefore for convexity along the KK direction it is most efficient to choose the KK charge a multiple of $N$.
	
	However, the opposite is true for winding. If we make $R$ extremely large, the dimension of operators with winding charge are dominated by the contribution from the $\mathrm{CFT}_{S^1}$ sector, which leads to a convex spectrum. For example, for the case of $R \gg N^{\frac{3}{2}}$ we see that the winding direction is convex from charge one. The index is therefore $I \sim \sqrt{N}$,
	which is much smaller than the spectral flow bound.
	
	Indeed, it is not possible to force the index to be larger than $\sqrt{N}$. This is because once we have a pure KK state as a convex direction, then for the other convex direction we are free to choose whatever KK charge we like. It will not contribute to the determinant since the first direction has no winding charge. We may then choose it to have winding charge one, and KK charge as large as necessary to ensure that the direction is convex. This leads to an index of $\sqrt{N}$.  
	
	\subsubsection*{$\mathrm{Case\;2}\;:\; R^2 \leq N$}
	
	In this case the largest $R$ can be is $\sqrt{N}$. For this we can choose a convex direction as having pure KK charge, starting at KK charge $\sqrt{N}$. As discussed above, we can then choose the other convex direction as having winding charge one, and any sufficiently large KK charge to ensure convexity. Therefore, in this case, it is not possible to force the index to be larger than $I \sim N^{\frac{1}{4}}$. Again, far from saturating the spectral flow bound. 
	
	\subsubsection*{Index for the strong version}
	
	If we consider the Strong Convexity Index Conjecture, then we require a sub-lattice which is itself composed of convex vectors. The sub-lattices considered above do not satisfy this requirement. For example, we can consider the case $R=1$. Then under the symmetries $U(1)_n \times U(1)_w$, the charge vectors $(1,0)$ and $(N,1)$ are convex, and act as a basis for a sub-lattice with index $I=1$. However, the strong version demands that also, say, $(0,1)$ must be convex, since it is in the sub-lattice. For $R=1$, this is not true, and so the basis $(1,0)$ and $(N,1)$ is not appropriate for the strong version. Indeed, allowing for general $R$, we require a minimal basis of $(N,0)$ and $(0,N)$, and therefore the index of the strong version saturates the spectral flow bound $I=N$.
	
	%%%%%%%%%%%%%%%%%%%%%%%%%%%%%%%%%%%%%%%%%%%%%%%
	\section{Higher dimensional CFTs}
	\label{sec:hdCFT}
	%%%%%%%%%%%%%%%%%%%%%%%%%%%%%%%%%%%%%%%%%%%%%%%
	
	In three or more dimensions, the proposal is that convexity cannot be parametrically delayed. In the case of multiple symmetries, it is natural to formulate this using the convexity index (\ref{index}), demanding that it cannot be made parametrically large. 
	
	In the two-dimensional case, we were able to delay convexity by introducing a discrete parameter $N$. In this section we study the possibility of doing something similar in higher dimensions. We focus on supersymmetric theories where we can consider BPS operators. Since BPS operators have a dimension proportional to their charge, they are natural states to (marginally) satisfy convexity. The first question we address is: Is it possible to have a parametrically large convexity index for BPS operators? By this we mean that we consider convex directions along BPS operators only, and calculate the index of the sub-lattice generated by these directions. We show, using explicit examples, that this index can be made parametrically large.
	
	This means that, at least in those examples, a bounded convexity index must utilise directions in charge space which are associated to non-BPS operators. It is difficult to determine such non-BPS convex directions, since they require loop calculations to test convexity. However, there is a way that a charge direction can be proven to not be convex.  We can consider a situation where the operators with charges smaller than $N$ are not BPS, while the charge $N$ operator is BPS. This would violate convexity: we could consider two integers $n_1$ and $n_2$ which sum to $N$, $n_1+n_2=N$, then
	\be
	\Delta\left(n_1 \bq\right) + \Delta\left(n_2 \bq\right) > \Delta\left(n_1 \bq\right)_{\mathrm{BPS}} + \Delta\left(n_2 \bq\right)_{\mathrm{BPS}} = \Delta\left(N \bq\right)_{\mathrm{BPS}} = \Delta\left(N \bq\right) \;,
	\label{bpsnonconvex}
	\ee
	where by $\Delta\left(n\right)_{\mathrm{BPS}}$ we denote the dimension of a BPS operator of charge $n$. Here we used that the operators of charges $n_1$ and $n_2$ are not BPS, while the operator of charge $N$ is BPS. We therefore study if, in the cases when the BPS convexity index can be made parametrically large, is it also possible to rule out using this method the non-BPS directions as convex. In the examples we study we find that this is not possible. Therefore, a check on convexity requires explicit calculations of the dimensions, which is beyond the scope of this work. 
	
	Note that the strong version of the Convexity Index Conjecture would typically directly imply having to go beyond the BPS spectrum, since typically BPS operators do not define a lattice. 
		
	%%%%%%%%%%%%%%%%%%%%%%%%%%%%%%%%%%%%%%%%%%%%%%%
	\subsection{The $XYZ$ model in three dimensions}
	\label{sec:xyzmod}
	%%%%%%%%%%%%%%%%%%%%%%%%%%%%%%%%%%%%%%%%%%%%%%%
	
	To illustrate the points above, we can consider a simple example: a three-dimensional ${\cal N}=2$ supersymmetric theory with three chiral superfields $X$, $Y$, $Z$ and a superpotential
	\be
	W = XYZ \;.
	\ee
	The theory flows to an interacting CFT in the infrared. It has three global $U(1)$ symmetries, $U(1)_R$, $U(1)_A$ and $U(1)_B$, under which the chiral fields transform as\footnote{We have rescaled the $U(1)_R$ charges so that they are integer.}
	\be
	\begin{array}{c|c|c|c}
	\mathrm{Operator} & U(1)_R & U(1)_A & U(1)_B \\
	\hline
	X & 1 & 1 & 1 \\
	Y & 1 & 0 & -2 \\
	Z & 1 & -1 & 1 \\	
	\end{array}
	\ee 
	The BPS operators are those which can obtain a vacuum expectation value along the moduli space, so they are $X^n$, $Y^n$, $Z^n$ for $n \in \mathbb{N}$. This theory has a BPS convexity index, so the index of the sub-lattice of BPS operators within the lattice of all operators, which is of order one. 
	
	We now gauge a $\mathbb{Z}_{2N}$ discrete subgroup of the global symmetry $U(1)_A$. This means that we project out all operators that are charged under this discrete symmetry. So, for example, $X^n$ and $Z^n$ are projected out unless $n$ is a multiple of $2N$. Let us now consider which operators are left in the spectrum. We will focus on bosonic operators; including also the fermionic operators leads to changes which are of order 1 and do not change the main results. We list these bosonic operators, noting which are BPS and which are not:
	 \be
	\begin{array}{c|c|c|c}
	\mathrm{Operator} & U(1)_R & U(1)_A & U(1)_B \\
	\hline
	X^{2N} \;(\mathrm{BPS}) & 2N & 2N & 2N \\
	Y \;(\mathrm{BPS})& 1 & 0 & -2 \\
	Z^{2N} \;(\mathrm{BPS}) & 2N & -2N & 2N \\	
	\hline
	XZ & 2 & 0 & 2 \\
	XYZ & 3 & 0 & 0 
	\end{array}
	\ee 
	We should first determine the lattice of charged bosonic operators. This is generated by the matrix of basis vector $T$ as
	\be
	T = \left( \begin{array}{ccc} 2 & 0 & 3 \\ 0 & 2N & 0 \\ 2 & 0 & 0 \end{array} \right) \;.
	\ee
	A linear transformation mapping this basis to the standard integer lattice is just given by $T^{-1}$. 
	Next we need to calculate the sub-lattice generated by the BPS operators, which is given by the basis vectors
	\be
	S = \left( \begin{array}{ccc} 1 & 2N & 2N \\ 0 & 2N & -2N \\ -2 & 2N & 2N \end{array} \right) \;.
	\ee
	Now we should act with the linear transform which mapped the lattice of charged operators to the integer lattice, and calculate the determinant, which gives the index
	\be
	I_{\mathrm{BPS}} = \left|\mathrm{det\;} \left(T^{-1} S\right) \right|^{\frac{1}{3}} = \left( 2N \right)^{\frac{1}{3}} \;.
	\ee
	We see that in this example, the sub-lattice of BPS operators has a parametrically large index within the lattice of operators. Therefore, we cannot have convexity with a bounded index using only BPS states. 
	
	As discussed at the beginning of this section, we could try to show an obstruction to convexity along the non-BPS operators by showing that at some parametrically large charge along a would-be convex direction there appears a BPS operator. However, it is clear that such an obstruction could never arise by gauging a discrete $\mathbb{Z}_N$ symmetry within a global $U(1)$ symmetry, since this projects out all operators, BPS or not, in that direction in charge space with charges $\leq N$ under this $U(1)$. So order-1 directions in charge space along non-BPS operators will never hit a BPS operator at some point. Determining convexity in this example therefore requires understanding the non-BPS physics. As mentioned above, including the fermions only leads to some order-1 shifts of the basis vectors, and leads to the same conclusions.

	%%%%%%%%%%%%%%%%%%%%%%%%%%%%%%%%%%%%%%%%%%%%%%%
	\subsection{Supersymmetric QCD in four dimensions}
	\label{sec:susyqcd}
	%%%%%%%%%%%%%%%%%%%%%%%%%%%%%%%%%%%%%%%%%%%%%%%
	
	In this section we give another example with a parametrically large BPS sub-lattice index. Another interesting aspect of this example is that we do not gauge a discrete subgroup of a $U(1)$ symmetry. We therefore avoid the general obstruction in the previous example to showing that some (non-BPS) directions in charge space are not convex by having a BPS operator at large charge along them.
	
	We will study two different ways to realise these properties in the same theory. The first is gauging a discrete $\mathbb{Z}_N$ global symmetry (rather than a discrete subgroup of a continuous one) for arbitrarily large $N$, which we discuss in section \ref{sec:gaugediscre}. The second is by tuning parameters in the theory, which we discuss in section \ref{sec:largeRmod}. 
	
	We consider the conformal window of four-dimensional $\mathcal{N}=1$ supersymmetric QCD (SQCD) \cite{Seiberg:1994pq}. The theory consists of an $SU(N_c)$ gauge group together with $N_f$ fundamental quarks $Q_i$ and $N_f$ antifundamental quarks $\tilde Q_i$. We denote the gaugino by $\lambda$ and the components of the quarks by $\phi_i,\psi_i$ and $\tilde{\phi}_i,\tilde{\psi}_i$. The anomaly-free continuous symmetry group is
	\begin{equation}
	SU(N_f)\times SU(N_f)\times U(1)_R\times U(1)_B \;.
	\end{equation}
	There is an additional classical chiral axial symmetry $U(1)_A$ which is anomalous and so is explicitly broken to a discrete $\mathbb{Z}_{2N_c}$ symmetry. Under these symmetries, the various fields have the following charges:
	\begin{center}
		\begin{tabular}{c|c c c c}
			& $SU\left(N_f\right)\times SU\left(N_f\right)$ & $U(1)_{R}$ & $U(1)_{B}$ & $\mathbb{Z}_{2N_c}$ \tabularnewline
			\hline 
			$\phi$ & $\left(N_f,1\right)$ & $1-\frac{N_c}{N_f}$ & $1$ & $1$ \tabularnewline
			$\psi$ & $\left(N_f,1\right)$ & $-\frac{N_c}{N_f}$ & $1$ & $0$ \tabularnewline 
			$\tilde{\phi}$ & $\left(1,\overline{N_f}\right)$ & $1-\frac{N_c}{N_f}$ & $-1$ & $1$ \tabularnewline
			$\tilde{\psi}$ & $\left(1,\overline{N_f}\right)$ & $-\frac{N_c}{N_f}$ & $-1$& $0$ \\
			$\lambda$ & $(1,1)$ & $1$ & $0$& $1$
		\end{tabular}
		\par\end{center}
	The R-charge is obtained by finding the non-anomalous combination of the classical R-symmetry and axial symmetry. 
	
	The operators above are not gauge-invariant, and so when we analyze their spectrum we must focus only on gauge-invariant combinations. There is an important subset of gauge-invariant operators which are also BPS, given by the mesons $M$ and baryons $B,\tilde B$:
	\begin{center}
		\begin{tabular}{c|c c c}
			& $SU\left(N_f\right)\times SU\left(N_f\right)$ & $U(1)_{R}$ & $U(1)_{B}$ \tabularnewline
			\hline 
			$M_{ij}=Q_i^a\tilde Q_j^a$ & $(N_f,\overline{N_f})$ & $2\frac{N_f-N_c}{N_f}$ & $0$ \tabularnewline
			$B_{[i_1...i_{N_c}]}=\epsilon_{a_1...a_{N_c}}Q_{i_1}^{a_1}...Q_{i_{N_c}}^{a_{N_c}}$ & $(\begin{pmatrix}N_f\\N_c\end{pmatrix},1)$ & $N_c\frac{N_f-N_c}{N_f}$ & $N_c$ \tabularnewline 
			$\tilde B_{[i_1...i_{N_c}]}=\epsilon_{a_1...a_{N_c}}\tilde Q_{i_1}^{a_1}...\tilde Q_{i_{N_c}}^{a_{N_c}}$ & $(1,\overline{\begin{pmatrix}N_f\\N_c\end{pmatrix}})$ & $N_c\frac{N_f-N_c}{N_f}$ & $-N_c$ \tabularnewline
		\end{tabular}
		\par\end{center}
	Here, $a$ indices denote gauge indices which we contract in gauge-invariant combinations, while $i$ indices denote flavor indices. 
	
	The dynamics of the theory depend on the values of $N_f,N_c$. In particular, there is a range in which the theory flows to an interacting CFT, called the conformal window. We will be interested in the Veneziano limit of $N_f,N_c\to \infty$ with $N_f/N_c$ kept constant; then the conformal window is given by the range $3/2\leq N_f/N_c\leq 3$. In particular, for $N_f$ close to $3N_c$, the fixed point becomes weakly-coupled.
	
	In the conformal window, the full quantum moduli space of the theory coincides with the classical one. The possible vevs are
	\begin{equation}
	Q=\begin{pmatrix}
	a_1 & & & \\
	& a_2 & & \\
	& & \ddots  & \\
	& & & a_{N_c} \\
	\\
	\\
	\end{pmatrix},\qquad 
	\tilde Q=\begin{pmatrix}
	\tilde{a}_1 & & & \\
	& \tilde{a}_2 & & \\
	& & \ddots & \\
	& & & \tilde{a}_{N_c} \\
	\\
	\\
	\end{pmatrix}
	\end{equation}
	where $|a_i|^2-|\tilde a_i|^2$ is independent of $i$. In terms of the gauge-invariant fields this is
	\begin{equation}
	M_{ii}=a_i\tilde a_i\;,\qquad i=1,...,N_c
	\end{equation}
	and 
	\begin{equation}
	B^{1...N_c}=a_1 a_2 ... a_{N_c},\qquad \tilde B^{1...N_c}=\tilde a_1 \tilde a_2 ... \tilde a_{N_c}\;,
	\end{equation}
	where all other elements not related to these by symmetry vanish. This allows us to read off the chiral ring, as the operators which can obtain an expectation value, and so the combinations of $M,B,\tilde B$ which are BPS; in particular, $M^n$ is BPS for all $n\in\mathbb{N}$, while $\det M$ is not BPS.
	
	We are interested in the sub-lattice generated by BPS states. We will focus on the $U(1)_R$ direction in charge space. The baryonic operators have parametrically large charges, by a factor of $N_c$, relative to the mesons, so we can focus on the meson operators. The operators, with charges under $U(1)_R$ and $\mathbb{Z}_{2N_c}$ are (with BPS operators denoted so):
	\be
	\begin{array}{c|c|c}
	\mathrm{Operator} & U(1)_R & \mathbb{Z}_{2N_c}  \\
	\hline
	\phi \tilde{\phi} \;(\mathrm{BPS}) & 2-\frac{2N_c}{N_f}  & 2  \\
	\phi \tilde{\psi}  & 1-\frac{2N_c}{N_f} & 1 \\
	\psi \tilde{\psi}   &-\frac{2N_c}{N_f} & 0 	
	\end{array}
	\label{mesonop}
	\ee

	%%%%%%%%%%%%%%%%%%%%%%%%%%%%%%%%%%%%%%%%%%%%%%%
	\subsubsection{Gauging the discrete $\mathbb{Z}_{2N_c}$ symmetry}
	\label{sec:gaugediscre}
	%%%%%%%%%%%%%%%%%%%%%%%%%%%%%%%%%%%%%%%%%%%%%%%

If we now gauge the $\mathbb{Z}_{2N_c}$ symmetry, we see from (\ref{mesonop}) that all BPS operators start with charge in the $U(1)_R$ direction of order $N_c$. They can be either the Baryons or the $\left(\phi \tilde{\phi} \right)^{N_c}$ mesons. The charge lattice for all operators starts with charges along $U(1)_R$ of order one, since we have the fermionic mesons $\psi \tilde{\psi}$. Therefore the BPS sub-lattice index grows with $N_c$, and can be made parametrically large.\footnote{More precisely, we should multiply all charges by $N_f$ so that all operators have integer charges. But this does not modify the index.}

To extract the power of $N_c$ in the index, we need to decide which lattice we consider. There are only two Abelian symmetries, $U(1)_R$ and $U(1)_B$, so we may consider this two-dimensional lattice. In which case the (normalised, as in (\ref{index})) index of the sub-lattice grows as $\sqrt{N_c}$. However, the full non-Abelian group include the important $SU(N_f)\times SU(N_f)$ factor. As discussed in section \ref{sec:lattind}, our proposal in that case is to include one Cartan sub-algebra representative from each non-Abelian factor, so that we have overall four $U(1)$ factors. The normalised index therefore behaves as 
\be
I \sim N_c^{\frac{1}{4}} \;.	
\ee
	Another aspect of this example that we are interested in is whether we can find non-convex directions using BPS states. So, in this case, we would like to find operators whose direction in charge space aligns with the remaining BPS mesonic operators $M^N_{ii}$. This would guarantee non-convexity.
	
%	Consider the SQCD conformal window. For R-charges of order 1 (as opposed to order $N_c$), the operator $m_n=(\tilde\phi_i\phi_j)^n$ is the only operator which saturates the BPS bound, where we use the $n^{\mathrm{th}}$ power to denote a symmetric product of $n$ such operators. The operator $m_n$ is thus in the representation $r_n=\text{sym}^n(r_1)$, where $r_1$ denotes the representation of $m_1$. We would thus like to project out $m_n$ for all $n<N$ with $N$ arbitrarily large. However, we must make sure that there remains some operator $\mathcal{O}$ with order-1 charges which is in the same direction in charge space as $m_N$. More precisely, this means that $\mathcal{O}$ must be in a representation $r_0$ such that $\text{sym}^k(r_0)$ includes the representation $r_N$. If this can be achieved, then the direction in charge space along $r_0$ will not be convex.  
%		
%	We will project out $m_n$ for $n<N$ by gauging the discrete $\mathbb{Z}_{2N_c}$ symmetry. The operator $m_n$ is charged under this symmetry for $n\neq 0\mod N_c$, and so in particular we project out all $m_n$'s for $n<N_c$. Now we must look for an operator $\mathcal{O}$ in the same direction as $m_{N_c}$ which is not projected out. 
	
	However, it turns out that such an operator does not exist. For example, possible choices of an operator are
	\begin{equation}
	\psi_i\lambda^{2}\tilde{\psi},\;\;\phi\lambda\tilde{\psi}\;,
	\end{equation}
	which have the same global charges as $M_{ii}$, and so are in the same direction in charge-space as $M^N_{ii}$. However, these operators are also charged under the discrete symmetry and so they are projected out. Alternatively, one can try to use operators of the form
	\begin{equation}
	\psi_{i}\tilde{\psi}_{j},\;\;\phi_{i}\overline{\lambda}^{2}\tilde{\phi}_{j}\;,
	\end{equation}
	which are not charged under the discrete symmetry and have the same $SU(N_f)\times SU(N_f)$ representation as $M_{ii}$, but they cannot be made to have the same R-charge, and so they are not in the same direction in charge-space. Products of these operators also cannot simultaneously be brought to the same direction in charge-space while not being projected out. 
	
	As a result, while we can project out all BPS operators up to some arbitrarily large R-charge, we cannot find an operator in the same direction in charge-space which is not projected out. This means that along the direction in charge space which passes through the BPS operator that is not projected out, convexity is maintained (exactly marginally).

	%%%%%%%%%%%%%%%%%%%%%%%%%%%%%%%%%%%%%%%%%%%%%%%
	\subsubsection{Large R-charge}
	\label{sec:largeRmod}
	%%%%%%%%%%%%%%%%%%%%%%%%%%%%%%%%%%%%%%%%%%%%%%%
	
	We now discuss an alternative method to find a non-convex direction in charge space. Assume that there exists a BPS operator $\mathcal{M}$ whose R-charge is arbitrarily close to $1$. Expanding $\mathcal{M}$ in components as
	\begin{equation}
	\mathcal{M}=\phi+\theta \psi +\theta^2  F\;,
	\end{equation}
	the corresponding R-charges are
	\begin{equation}
	r_{\phi}=1+\epsilon,\qquad r_{\psi}=\epsilon,
	\end{equation} 
	where $|\epsilon|\ll 1$. The convexity conjecture requires normalizing $U(1)$ charges such that the smallest charge is $1$. Denoting the renormalized charges by $r'$, the renormalized charges are
	\begin{equation}
	r'_{\phi}=1+1/\epsilon,\qquad r'_{\psi}=1.
	\end{equation} 
	Now it is clear that in terms of renormalized charges, there is a BPS operator $\phi$ with arbitrarily large charge. If in addition there are no other BPS operators in this direction in charge-space (while at least some operators exist in this direction with order-1 charges), then this provides a non-convex direction in charge space.
	
	Let us show how this method can be applied to the SQCD conformal window. The only BPS operators with order-one R-charges are products of the mesons $M$. Focusing on a single meson $M$, its R-charge is $R_M=2\frac{N_f-N_c}{N_f}$, and it can be brought arbitrarily close to 1 by taking $N_c/N_f$ arbitrarily close to $1/2$ (note that these values are still inside the conformal window). The normalised $U(1)_R$ charges of all the scalar mesons are now parametrically large. Since only the scalar mesons are BPS, this again gives BPS operators at parametrically large charge. 
	
	We can now look again for an operator with order-one charges which is in the same direction in charge-space as the high-charge BPS operators. However, again, we do not find any such operators. So, like in the other cases, convexity depends on the convexity of the non-BPS operators.  
			
	%%%%%%%%%%%%%%%%%%%%%%%%%%%%%%%%%%%%%%%%%%%%%%%
	\section{Summary}
	\label{sec:summary}
	%%%%%%%%%%%%%%%%%%%%%%%%%%%%%%%%%%%%%%%%%%%%%%%
	
	In this paper we studied aspects of the convexity conjecture of \cite{Aharony:2021mpc}, which proposes that charged operators in CFTs have a convex spectrum along some directions in charge space. We developed a sharp formulation of it in the case of multiple Abelian (and non-Abelian) symmetries with respect to at what charge convexity should appear. We introduced the convexity index, which is the index of the sub-lattice of operators generated by a convex basis of vectors in charge space. We then proposed that this index cannot be made parametrically large in three dimensions or more. 
	
	In two dimension the convexity index can be made parametrically large, which we prove with an explicit example. On the other hand, we also prove that it is bounded in a precise way by the levels of the current algebra associated to the global symmetries. In this sense, the two-dimensional case can be thought of as a setup where the conjecture needs to be modified slightly to account for the level associated to the $U(1)$ global symmetry, and once this is accounted for, it can be rigorously proven.
	
	In three and higher dimensions, we studied examples. We considered supersymmetric theories and studied whether BPS operators are sufficient to show the convexity conjecture. We showed that there are theories where the convex vectors along BPS operators generate a sub-lattice with a parametrically large index. This shows that the convexity conjecture, at least in its current formulation, cannot be satisfied by looking at the BPS spectrum only. It must involve the non-BPS operator spectrum. 
	
	It is possible that perhaps a modification of the conjecture in higher dimensions, similar to the one required in two dimensions, would account for the parametrically large BPS sub-lattice index. It would be interesting to develop this. However, note that in two-dimensional CFTs there is an integer level that appears naturally in a modified conjecture. We are not aware of such a natural integral quantity in higher dimensions.
	
	This integer quantity has a nice holographic dual as the Chern-Simons level of the dual gauge field. Therefore, our results are also a proof of the Positive Binding Conjecture of \cite{Aharony:2021mpc} in three-dimensional gravity, up to the modification that the charge of the particle with positive self-binding energy is bounded by the Chern-Simons level. This conjecture has been argued in \cite{Aharony:2021mpc} to be the natural formulation of the Weak Gravity Conjecture in AdS. Indeed, our methodology is closely related to the analyses of the Weak Gravity Conjecture in \cite{Montero:2016tif} and \cite{Heidenreich:2016aqi}. 
	
	We also proposed a strong version of the conjecture, which demands that every point in the sub-lattice itself defines a convex vector. This is motivated by studies of lattices of charged states in flat space \cite{Heidenreich:2016aqi,Heidenreich:2019zkl}. With regards to the results of this work, the evidence for the weak and strong versions is the same.

	\vskip 20pt
	\noindent {\bf Acknowledgements:} We thank Ofer Aharony, Ralph Blumenhagen and Cumrun Vafa for very useful discussions. The work of EP was supported by the Israel Science Foundation (grant No. 741/20) and by the German Research Foundation through a German-Israeli Project Cooperation (DIP) grant "Holography and the Swampland". A.S. is supported by an Israel Science Foundation center for excellence grant (grant number 2289/18), by the Minerva foundation with funding from the Federal German Ministry for Education and Research, by grant no. 2018068 from the United States-Israel Binational Science Foundation (BSF), by the German Research Foundation through a German-Israeli Project Cooperation (DIP) grant "Holography and the Swampland", and by a research  grant from Martin Eisenstein.
	
	\appendix

	\bibliographystyle{ssg}
	\bibliography{susyswamp.bib}  
	%$\bibliographystyle{custom1}
\end{document}